\documentclass[3p, preprint]{elsarticle}
\begin{document}

\title{ How to address
cellular heterogeneity by distribution biology}
\date{15th December 2016}
\author[lcsb]{Niko~Komin}
\ead{niko.komin@uni.lu}
\author[lcsb,ucsd]{Alexander~Skupin}
\ead{alexander.skupin@uni.lu}

\address[lcsb]{Luxembourg Centre for Systems Biomedicine, University of Luxembourg, Belval, Luxembourg}
\address[ucsd]{University California San Diego, La Jolla, USA}

\begin{abstract}
Cellular heterogeneity is an immanent property of biological systems that covers very different aspects of life ranging from genetic diversity to cell-to-cell variability driven by stochastic molecular interactions, and noise induced cell differentiation. Here, we review recent developments in characterizing cellular heterogeneity by distributions and argue that understanding multicellular life requires the analysis of heterogeneity dynamics at single cell resolution by integrative approaches that combine methods from non-equilibrium statistical physics, information theory and omics biology. 
\end{abstract}

\maketitle

\section*{Introduction}
Life is heterogeneous -- at nearly all biological scales and
levels~\cite{Altschuler2010}. Maybe the most obvious heterogeneity can
be observed when looking at the different species that evolution has
created. But heterogeneity is much further spread in living systems
where individuals within one species exhibit unique properties and
even cells of the same cell type within the same (multicellular)
organism can possess a wide range of divergent characteristics.
 
An essential function of heterogeneity is to ensure robustness of a
biological system in fluctuating environments~\cite{Ackermann2015Functional}. Thus, the
spread of phenotypic traits within a population of a species allows
for a broader niche in which the population as a whole can survive and
adapt to different conditions including competition for
resources. This central mechanism of life has led to the general
perception that "Nothing in Biology Makes Sense Except in the Light of
Evolution" as stated by Theodosius Dobzhansky in his landmark essay in
1973~\cite{Dobzhansky1973Nothing}.  While Dobzhansky was mainly
focusing on the diversity of different organisms and criticizing
creationism, his conclusion may be also instructive to address the
currently urgent open question on multicellularity as the foundation
of multicellular organisms and microorganism colonies.

A central problem in multicellularity arises as to how an organism
can give rise to all desired different cell types originating from the
same genome in an coordinated manner~\cite{Huang2010Cell,
  Huang2009Nongenetic}. Considering the concept of evolution as a
general underlying mechanism of life with its two major components,
mutation and the interplay of induction and selection, we may scale
Dobzhansky's concept down to the level of cell populations. Thereby,
mutations can be generalized to immanent heterogeneity and the
interplay between induction and selection represents intra- and
intercellular signaling. Hence, understanding (multicellular) life
across its different scales relies on investigating {\it cellular
  nanoevolution} - that is the dynamics of cellular heterogeneity.

\begin{figure}[htbp]
\begin{center}
\includegraphics[height=0.5\textwidth,angle=270]{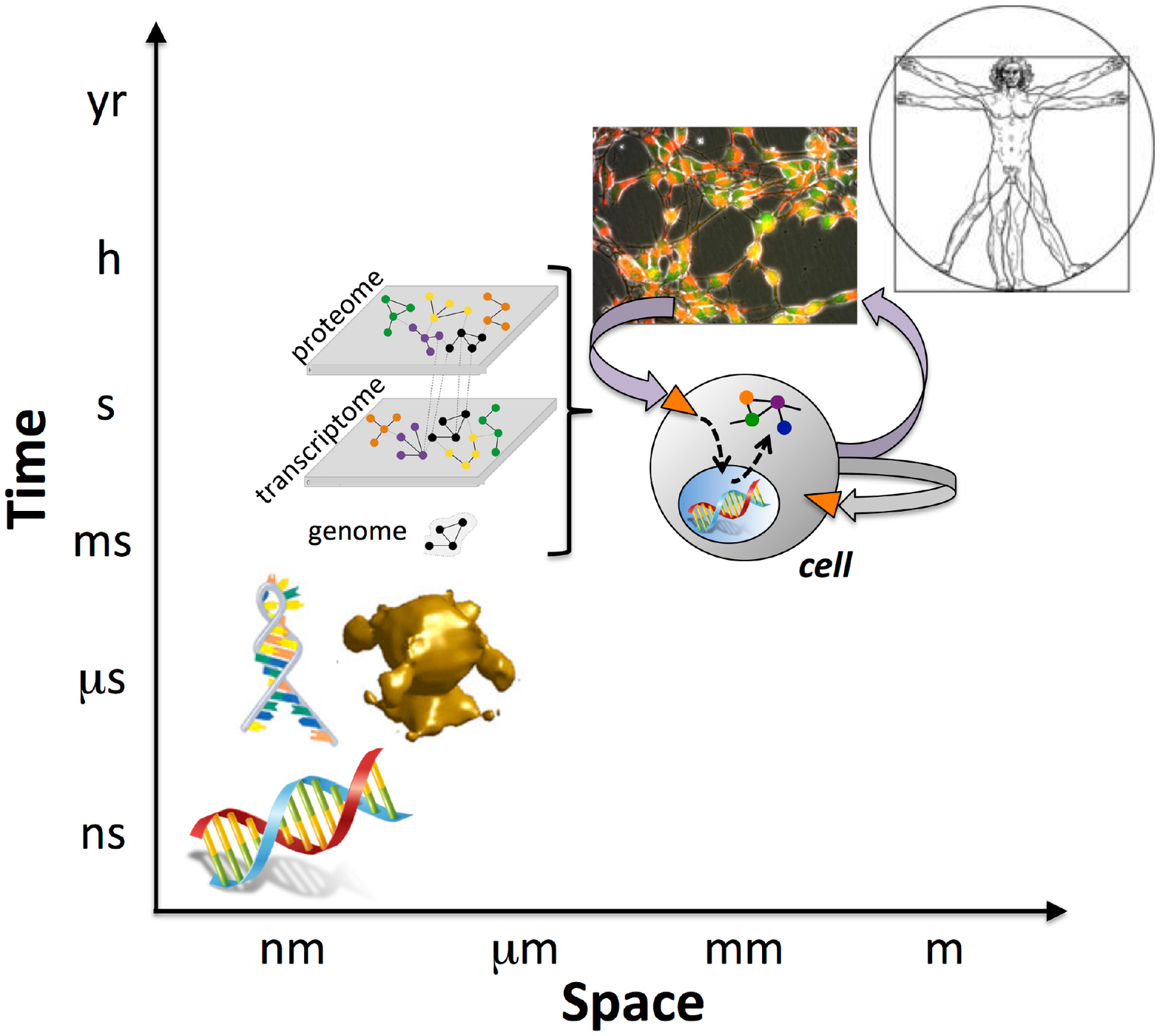}
\caption{{\bf Multiscale organization of life.} Considering the
  different scales of multicellular cell fate from noisy gene
  expression to intercellular signaling enables adaptation and
  reliable morphogenesis.(Some figure elements are taken from ~\cite{Farris2015,Putney1993} and permissions are currently pending.)}
\label{fig:multiscale}
\end{center}
\end{figure}

The origin of cellular heterogeneity (even within clonal populations)
is the multiscale organization of life as depicted in
Fig.~\ref{fig:multiscale}. On the smallest relevant scales, the
stochastic nature of molecular interactions induces individual
transcription profiles that are subsequently translated into
heterogeneous cellular phenotypes. These rather randomly induced
phenotypes are subsequently instructed and selected on the level of
the cell population by intercellular signaling or cell-cell interactions leading to the coordinated generation
of tissues, organs and organisms. This underlying
noise-driven cellular heterogeneity is the mechanism to balance
cellular robustness and adaptability~\cite{Garcia-Ojalvo2012}.

Until recently, cellular heterogeneity was only accessible on rather
low dimensional readouts such as specific protein abundance by
antibody staining or since more recently by flux cytometry
analysis. These limited investigations hindered a more systematic
approach to dissect underlying mechanisms. But with the recent
developments of several single cell analysis methods
\cite{Sato2010,Chattopadhyay2014,Patel2014,Shalek2014,Junker2014,Macosko2015}, we are now able to characterize cellular heterogeneity in great
detail. Despite these advancements, a systematic approach how to
interpret and use the resulting high-dimensional data for identifying
biological principles of multicellular life is still
lacking. Currently, methods from statistical physics are intensively
discussed to be exploited but the underlying non-equilibrium dynamics
and biological complexity of intra- and intercellular interactions
make a direct application
difficult~\cite{Garcia-Ojalvo2012,Pujadas2012,MacArthur2013}.

While previous reviews summarize potential functions of cellular
heterogeneity from an experimental perspective
~\cite{Ackermann2015Functional, Dueck2016Variation}, we will review
here more recent developments and how mathematical modeling and
analysis can be used to reveal general functions and guiding
principles from the central descriptor of heterogeneity -- the (phenotype) trait distribution.

\section*{Characterization of cellular heterogeneity}
Heterogeneity of cells can have different properties
(Fig.~\ref{fig:heterogeneity}) and lead to multiple benefits for
the
population~\cite{Ackermann2015Functional,Martins2015Microbial,Dueck2016Variation}:
it ensures robustness of the biological system to fluctuating
environments  and
allows the population to adapt to a wider variety of environments (bet
hedging); binary decisions on a cellular
scale can yield a fractional or dose-dependent response; rare
individuals can coordinate population behavior by emitting local
signals; subpopulations can be primed for multiple cell fates;
transmitted and encoded information can be more complex.

\subsection*{Direct phenotypic heterogeneity}
Cell heterogeneity is observed relatively easily when it directly affects
phenotypic traits such as size distributions (e.g. axon diameter
variability~\cite{Buzsaki2014Logdynamic}) or cell survival times or
cell division times~\cite{Gascoigne2008Cancer, Spencer2009Nongenetic}.
The 
main driver of heterogeneity can be an underlying genetic
variability as exemplified in Fig.~\ref{fig:heterogeneity}A by the
biomass production of yeast strains. The observed trait variability
within the two strains YO490 and YO512 is much smaller compared to the
distribution of the progenies originating from a cross of these two
parental strains. The non-trivial effects of the genetic recombination
and the genome-environment interaction lead to a wide range of
biomass production for the progenies that is not simply between the
parental strains but exhibits more extreme traits.

But cellular heterogeneity does also occur independent of genetic
diversity such as in clonal populations due to the multiscale
organization depicted in Fig.~\ref{fig:multiscale}. Here, the
stochastic nature of molecular interaction induces random
transcription profiles that are subsequently modified in an
environment dependent manner including
signaling between cells. An related and medically important example of such a coordinated heterogeneity is the
epithelial-to-mesenchymal transition (EMT) and its counterpart MET as
a mechanism for metastasis~\cite{Thiery2009}. The underlying mechanism
is that a cancerous epithelial cell can transdifferentiate into a
mobile mesenchymal cells that is subsequently leaving the tumor, can
travel with the blood stream to distinct tissue sites where it
transforms back into an epithelial cell and initiating a secondary
tumor. The underlying mechanism can also be studied in model cell
lines like clonal HMLER cells where e.g. mesenchymal cell identity can
be determined by flux cytometry using cell surface markers such as CD44
~\cite{Grosse-Wilde2015}. The HMLER population exhibits a heterogeneous
steady-state distribution as shown in Fig.~\ref{fig:heterogeneity}B
where a unimodal phenotype distribution (blue) generated by cell
sorting is relaxing towards a stable bimodal distribution (gray)
that represents a mixture of epithelial (low CD44) and
mesenchymal (high CD44) cells. Computational modeling has predicted
that partial EMT instead of complete EMT is associated with tumor
progression~\cite{Jolly2016Stability} and is currently under experimental investigation. Furthermore, EMT has been recently used
as a show case that the dynamics of gene circuits is mainly determined
by their topology and not by specific parameters that may explain the
stable bimodal trait distribution~\cite{Huang2016Interrogating}. 



\subsection*{Dynamical heterogeneity}
\begin{figure*}[]
\begin{center}
\includegraphics[height=0.99\textwidth,angle=270]{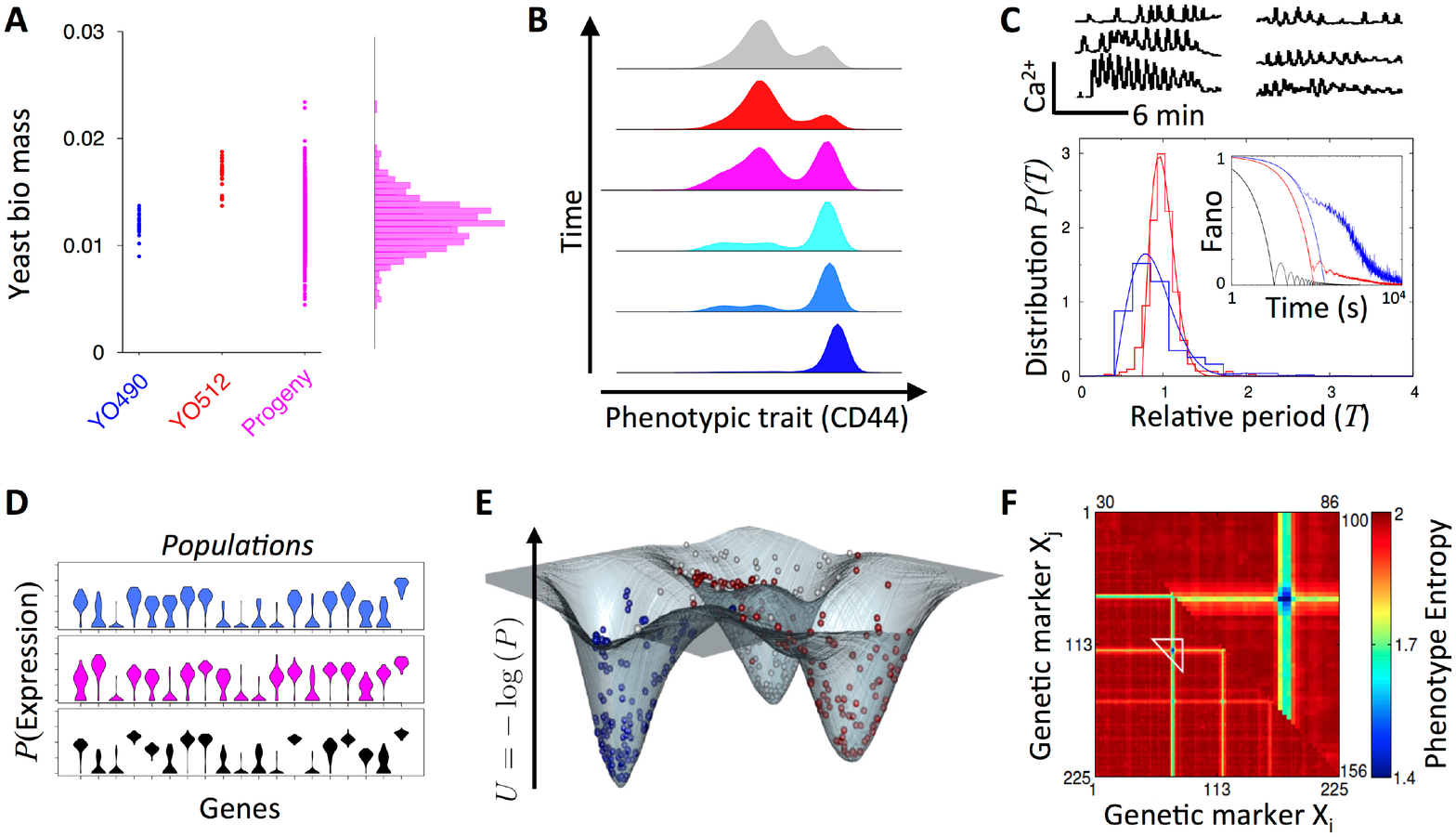}
\caption{{\bf Different aspects of heterogeneity.} {\bf A:} The
  biomass production of two different yeast strains and progenies from a
  cross of these exhibit different variability where the heterogeneity of the
  progenies is increased due to the genetic variability~\cite{TAN2013}.
  {\bf B:} Heterogeneity of a cell surface marker (CD44) in a clonal
  population of HMLER cells~\cite{Grosse-Wilde2015} characterized by
  flux cytometry. After sorting for mesenchymal cells (blue bottom) the
  cell population relaxes towards a stable mixture of epithelial and
  mesenchymal cells (gray) by mesenchymal-to-epithelial transition over
  6 weeks. 
   {\bf C:} Dynamic heterogeneity in cell signaling
  illustrated by calcium imaging within astrocytes exhibits random
  spiking behavior~\cite{Skupin2008How}. Thus, spikes occur randomly
  with the shown probability density for astrocytes (blue) and HEK
  cells (red). The stochastic process can be further characterized by
  the Fano factor (inlet)~\cite{Skupin2010Statistical}. 
   {\bf D:} Recent developments in single cell analysis methods allow for
  characterization of single cell transcription profiles as shown here
  for blood cell development from stem cells (black) to erythrocytes
  (magenta) and myeloid cells (blue) that exhibit distinct gene
  expression signatures~\cite{Mojtahedi2016Cell}.
  {\bf E:} To
  characterize differentiation dynamics, single cell transcriptomics
  data can be combined with dynamic systems theory to identify
  transitions between distinct cell state attractors as shown here by
  an inferred epigenetic landscape based on a PCA analysis of blood
  cell development where gray spheres correspond to stem cells, blue
  to myeloid cells and red to erythrocytes.  
  {\bf F:} Genetic
  heterogeneity such as generated by yeast crosses (A) can be used to
  identify genetic basis of phenotypic traits by statistical and
  information theory based methods~\cite{Ignac2014}.}
\label{fig:heterogeneity}
\end{center}
\end{figure*}

As illustrated by the example of EMT (Fig.~\ref{fig:heterogeneity}B), dynamics is essential for
cellular heterogeneity development.  
But also intracellular dynamics itself can exhibit large variability. The
maybe best studied stochastic dynamics in biology is the firing
pattern of neurons~\cite{Buzsaki2014Logdynamic}. The two main drivers of
the random dynamics are (i) the channel noise that originates from the
stochastic molecular interactions and (ii) the plethora of synaptic
connections each neuron has to neighboring cells. Theoretical and
computational approaches during the last decades have revealed the
rich dynamical spectrum of noise-driven neuronal dynamics where
non-linearities lead to non-trivial effects like stochastic and
coherent resonance in excitable media~\cite{Lindner2004}.

More recently, we have described similar characteristics in
Ca$^{\rm 2+}$ oscillations~\cite{Skupin2008How} of non-electrically excitable
cells. Ca$^{\rm 2+}$ as a central messenger in eukaryotic cells
transmits external signals into the cell by transient increases of the
cytosolic Ca$^{\rm 2+}$ concentration as those shown in
Fig.~\ref{fig:heterogeneity}C. While these transients have been
referred to as oscillations, we have demonstrated that they occur
randomly with a well defined probability density function $P(T)$ shown
in Fig.~\ref{fig:heterogeneity}C for astrocytes (blue) and HEK cells
(red). We combined statistical analysis~\cite{Skupin2010Statistical}
with multiscale modeling~\cite{Skupin2010Calcium} to demonstrate that
experimental observations are consistent with an hierarchical
signaling system where molecular noise of individual channels are
carried onto the level of the cell by diffusion mediated coupling of
release channels. Interestingly, the cellular heterogeneity exhibits
cell-type and pathway specific signatures (probability distributions
$P$) that are also encoded in the signal variability~\cite{Thurley2014}.

On a slower temporal scale, the circadian rhythm between sleep and activity of mammals is
generated by a number of neurons in the suprachiasmatic nucleus which
show phases of spontaneous firing alternating with quiet phases.
These sleep-wake cycles show periods distributed over a wide range of
durations when dispersed in a culture. Coupled in the tissue however,
they generate a rather precise rhythm in synchrony with an external
light stimulus~\cite{Honma2004Diversity}. For a model of coupled
biochemical clocks, subjected to periodic forcing we showed that an
intermediate value of period dispersion actually augments the
entrainability to the stimulus~\cite{Komin2011Synchronization}. In a
similar context an intermediate dispersion in the glutamate/orexin
threshold brings the sleep-wake cycle close to the optimal value~\cite{Patriarca2015Constructive}.


\subsection*{Cellular omics heterogeneity}
While the examples on cellular heterogeneity revised above are based
on low-dimensional observations, recent developments in single cell
analysis methods allow now for characterizing cellular heterogeneity
on the genomics and transcriptomics level~\cite{Macosko2015,Saadatpour2015,
  Shalek2013, Macaulay2015}. Characterizing
the resulting high-dimensional information by physiological means is essential
for the identification of biological mechanisms.  The set of measured
genes can be described as a vector in a gene expression space and
cluster analysis is commonly used to group
genes together by correlation and deducing functional proximity~\cite{Eisen1998Cluster}. Other
complexity reducing methods based on for example principle component analysis (PCA)~\cite{Korem2015Geometry},
stochastic neighboring embedding (tSNE)~\cite{Zeisel2015Cell} or diffusion maps including
pseudo-time ordering represent useful tools to
investigate correlative interactions in the high dimensional data~\cite{Angerer2016Destiny}. 

Despite the detailed characterization of cellular heterogeneity, an
integrative approach how to use the resulting high-dimensional data to
understand principles of multicellularity is missing. Recently, we
applied dynamical system theory to single cell transcription data to
investigate blood cell differentiation~\cite{Mojtahedi2016Cell}. By
treating hematopoietic stem cells with either EPO or IL-3/GM-CSF we
could induce differentiation into erythrocytes (red
blood cells)  and myeloid (bone marrow) cells,
respectively. Using single cell transcription analysis, we obtained
distinct gene expression profiles a subset of which is shown in
Fig.~\ref{fig:heterogeneity}D for stem cells (black), the erythroid
(magenta) and myeloid (blue) populations. To investigate the
underlying differentiation dynamics we were analyzing correlations
between cells and genes in a time and treatment dependent manner. In
agreement with insights from dynamical systems theory we could
demonstrate that differentiation occurs by the destabilization of an
stem cell attractor and the subsequent drug-induced development of the
2 linage attractors as visualized in Fig.~\ref{fig:heterogeneity}E by
a data-inferred epigenetic landscape (stem cells shown in gray,
erythrocytes in red and myeloid cells in blue). Additional analysis
based on properties of imperfect bifurcations validated
experimentally that the differentiation dynamics exhibit
characteristics of critical transitions~\cite{Trefois2015Critical} including
transient increase of heterogeneity and critical slowing down.

In a similar approach to transcription data sets of Drosophila, criticality of a 
biological system was investigated. The developmental gap gene network in the fruit fly embryo is
most likely presented by two mutually repressive genes and this simple
genetic network can be tuned to a critical point where one of the
eigenvalues vanishes. If spatially coupled, the resulting
system exhibits slow dynamics, strong local (anti-)correlations, and
long-ranged correlations in space, signs of criticality which can also
be validated quantitatively in the experimental
data~\cite{Krotov2014Morphogenesis}.

\subsection*{Use of genetic heterogeneity to identify molecular mechanisms}
The scope of the studies mentioned so far was to dissect the origin and function of cellular heterogeneity of different natures by combining quantitative experimental methods with theoretical and computational approaches. But genetic cellular heterogeneity can also be exploited to identify molecular mechanisms of phenotypic traits. For these approaches the awesome power of yeast genetics is often used to introduce genetic heterogeneity in a controlled manner by generating a large number of progenies from two parental strains (Fig.~\ref{fig:heterogeneity}A). The phenotypic diversity can then be mapped to specific gene loci by statistical inference or information theory methods~\cite{Ignac2014}. 

Using such an approach to understand the underlying reason for different morphologies of yeast colonies, we could identify chromosomal copy number variation as a multicellular phenotype switch~\cite{TAN2013}. Recently, Rabinowitz and colleagues were analyzing metabolic traits across different yeast cultures and inferred the data into Michaelis-Menten relationships between enzyme, substrate and products. This systematic analysis revealed previously unknown cross-pathway regulations and demonstrated that substrate concentrations are the strongest driver of metabolic reactions~\cite{Hackett2016Systemslevel}.

While monocausal genetic factors can be identified rather robustly, many phenotypic traits including disease development often result from interactions of several genes. The corresponding combinatorial explosion has to be compensated by large genotype-phenotype data sets that can be analyzed by statistical methods. To address non-linear genetic interactions, we have recently investigated different data sets by information theory based methods~\cite{Ignac2014}. By applying an entropy approach to the phenotype distributions, we were able to identify genetic interactions in the sporulation efficiency of yeast~\cite{Gerke2009} shown in Fig.~\ref{fig:heterogeneity}F and introduced the general measure of interaction distance for the identification of genetic synergies.

\section*{Conclusion}

Cellular heterogeneity is an immanent property of biological systems
originating from the multiscale organization of life shown in
Fig.~\ref{fig:multiscale}. This mechanism of random induction and
context dependent adaptation (or selection) can be seen as the
establishment of a \textit{cellular nanoevolution} where cells perform
random walks within a general configuration space (the epigenetic
landscape) according to a generalized chemotaxis. This analogy emphasizes
that evolutionary strategies may represent a generic biological
principle and that
multicellularity has to consider the dynamics of cellular
heterogeneity.

To address this heterogeneity dynamics, we have to integrate the
distributions covering the variability at the different biological scales
and levels (Fig.~\ref{fig:heterogeneity}). While such an integration has already been successfully performed on low dimensional traits like predicting glycogen distributions \textit{in vitro} based on maximal entropy principles \cite{Kartal2011a} and developing a corresponding theoretical description by open chemical network theory \cite{rao2015glucans}, a generalization to heterogeneous~\cite{Lafuerza2013Effect} and high-dimensional systems is not trivial. A major complication arises thereby from the strong non-equilibrium character of life and the different nature of distributions that ranges from genetic heterogeneity~\cite{TAN2013}, noise induced cell-to-cell variability on the transcription and protein level~\cite{Mojtahedi2016Cell} as well as individual dynamic properties that may originate from specific subcellular arrangements~\cite{Skupin2010Calcium}. This complexity renders a direct application of distribution based approaches from mathematics and statistical physics complicated. 

Nevertheless, distributions provide a promising perspective to tackle cellular heterogeneity and its role in multicellular life because they can bridge between biological complexity and methods from non-equilibrium statistical physics~\cite{Garcia-Ojalvo2012} and information theory~\cite{Voliotis2014Information}. Moreover, the recent experimental innovations in single cell analysis approaches enable now the generation of comprehensive data sets needed for the successful development of a \textit{distribution biology} framework that goes beyond the established network biology. Thereby the biological complexity may also trigger further theory developments in mathematics and physics indicating that, in accordance with Dobzhansky, "seen in the light of evolution, biology
is, perhaps, intellectually the most satisfying and inspiring science"~\cite{Dobzhansky1973Nothing}.

\section*{References}
\bibliographystyle{unsrt}
\bibliography{bibliography_corrected}
\pagebreak

\pagebreak

\end{document}